\documentclass[11pt,a4]{article}
\usepackage{latexsym,amssymb}

\begin{document}
\newcommand{\2}{\vspace{0.2 cm}}

\newcommand{\qed}{\hfill$\Box$}
\newcommand{\pf}{{\bf Proof: }}
\newtheorem{theorem}{Theorem}[section]
\newtheorem{lemma}[theorem]{Lemma}
\newcommand{\beq}{\begin{equation}}
\newcommand{\eeq}{\end{equation}}

\title{On Complexity of Minimum Leaf Out-Branching Problem}

\author
{Peter Dankelmann\thanks{School of Mathematical Sciences, University
of KwaZulu-Natal, Durban 4041, South Africa,
dankelma@ukzn.ac.za}\and Gregory Gutin\thanks{Corresponding author.
Department of Computer Science, Royal Holloway University of London,
Egham, Surrey TW20 0EX, UK, gutin@cs.rhul.ac.uk
} \and Eun Jung
Kim\thanks{Department of Computer Science, Royal Holloway University
of London, Egham, Surrey TW20 0EX, UK, E.J.Kim@cs.rhul.ac.uk} }

\maketitle

\begin{abstract}
\noindent Given a digraph $D$, the Minimum Leaf Out-Branching
problem (MinLOB) is the problem of finding in $D$ an out-branching
with the minimum possible number of leaves, i.e., vertices of
out-degree 0. Gutin, Razgon and Kim (2008) proved that MinLOB is
polynomial time solvable for acyclic digraphs which are exactly the
digraphs of directed path-width (DAG-width, directed tree-width, 
respectively) $0$. We investigate how
much one can extend this polynomiality result. We prove that already for
digraphs of directed path-width (directed tree-width, DAG-width,
respectively) $1$, MinLOB is NP-hard. On the other hand, we show that
for digraphs of restricted directed tree-width (directed path-width,
DAG-width, respectively) and a fixed integer $k$, the problem of
checking whether there is an out-branching with at most $k$ leaves
is polynomial time solvable.
\end{abstract}

\section{Introduction}
A digraph $T$ is an {\em out-tree} if
$T$ is an oriented tree with only one vertex $s$ of in-degree zero
(called {\em the root}). The vertices of $T$ of out-degree zero are
called {\em leaves} and all other vertices of $T$ are called {\em
nonleaves}. The vertex of in-degree zero is called the {\em root} of
$T$ and all vertices of out-degree at least 2 are called {\em
branching vertices}. If an out-tree $T$ is a spanning subgraph of
a digraph $D$, i.e.\ 
$V(T)=V(D)$, then $T$ is called an {\em out-branching} of $D$.

Given a digraph $D$, the {\em Minimum Leaf Out-Branching} problem
({\em MinLOB}) is the problem of finding in $D$ an {out-branching}
with the minimum possible number of leaves. Notice that not every
digraph $D$ has an out-branching. It is not difficult to see that
$D$ has an out-branching  if and only if $D$ has just one strong
connectivity component without incoming arcs \cite{bang2000}. Since
the last condition can be checked in linear time \cite{bang2000}, we
may often assume that a digraph $D$ has an out-branching.

The MinLOB problem on acyclic digraphs has applications in the area
of database systems, see the patent \cite{demers2000}, where a
heuristic to solve the MinLOB problem on acyclic digraphs was
suggested. Gutin, Razgon and Kim \cite{gutinAAIM08} showed that the
MinLOB problem for acyclic digraphs is, in fact, polynomial time
solvable. Since MinLOB extends the Hamilton path problem, MinLOB for
all digraphs is NP-hard, but standard dynamic programming techniques
allow one to have a polynomial time algorithm for digraphs whose
underlying graph is of bounded tree-width \cite{gutinAAIM08}.

In this paper we investigate how much we can extend the
polynomiality result for acyclic digraphs. Notice that acyclic
digraphs are the digraphs of directed path-width (directed
tree-width, DAG-width, respectively) 0. We prove that already for
digraphs of directed path-width (directed tree-width, DAG-width,
respectively) 1, MinLOB is NP-hard. This is in sharp contrast to the fact
that the Hamilton path problem (the most important special case of
MinLOB) is polynomial time solvable for digraphs of bounded directed
path-width (directed tree-width, DAG-width, respectively). This
fact follows from Theorem \ref{dtwth} and the inequalities on the
width parameters used in the proof of Theorem \ref{posth}.

On the other hand, we show that for digraphs of bounded directed
tree-width (directed path-width, DAG-width, respectively) and a
fixed integer $k$, the problem of checking whether there is an
out-branching with at most $k$ leaves is polynomial time solvable.

We consider directed path-width, directed tree-width and DAG-width
as they appear to be the most studied directed width parameters, but
our results hold for other width parameters such as elimination
width and Kelly-width \cite{hunterSODA07} (the results for
Kelly-width have to be modified by 1 taking into consideration that
Kelly-width equals elimination-width plus 1).

\section{Three Directed Decompositions}

DAG-width was introduced independently by Berwanger et al.
\cite{berwangerSTACS06} and Obdrzalek \cite{obdrzalekSODA06}. A {\em
DAG-decomposition (DAGD)} of a digraph $D$ is a pair $(H,\chi)$
where $H$ is an acyclic digraph and $\chi=\{W_h:\ h\in V(H)\}$ is a
family of subsets (called {\em bags}) of $V(D)$ satisfying the
following three properties: (a) $V(D)=\bigcup_{h\in V(H)}W_h$, (b)
if $u\in W_{h_1},\ v\in W_{h_2}$ and $(u,v)\in A(D)$, then there is
a directed $(h_1,h_2)$-path in $H$ (it is possible that $h_1=h_2$),
and (c) for all $h,h',h''\in V(H)$, if $h'$ lies on a directed path
from $h$ to $h''$, then $W_h\cap W_{h''} \subseteq W_{h'}$. The {\em
width} of a DAGD $(H,\chi)$ is $\max_{h\in V(H)}|W_h|-1$. The {\em
DAG-width} of a digraph $D$ (${\rm dagw}(D)$) is the minimum width
over all possible DAGDs of $D$.

A {\em directed path decomposition (DPD)} \cite{barat} is a special
case of DAGD when $H$ is a directed path. The {\em directed
path-width} of a digraph $D$ (${\rm dpw}(D)$) is defined as the
DAG-width above, but DAGDs are replaced by DPDs.

Directed tree-width was introduced by  Johnson, Robertson,
Seymour and Thomas \cite{johnsonJCT82}. A set $S \subseteq V(D)-Z$
is $Z$-{\em normal} if every directed walk that leaves and again
enters $S$ must traverse a vertex of $Z$. For vertices $r,r'$ of 
an out-tree $T$ we write
$r\le r'$ if there is a path from $r$ to $r'$ or $r=r'.$ An {\em
arboreal decomposition} of a digraph $D$ is a triple $(R,X,W)$,
where $R$ is an out-tree (not a subgraph of $D$), $X=\{X_e:\ e\in
A(R)\}$ and $W=\{W_r:\ r\in V(R)\}$ are families of sets of vertices of $D$ that
satisfy two conditions: (1) $\{W_r:r \in V(R)\}$ is a partition of
$V(D)$ into nonempty sets, and (2) for each $e=(r',r'')\in A(R)$
the set $\bigcup \{W_r:\ r\in V(R), r\ge r''\}$ is $X_e$-normal. The
{\em width} of $(R,X,W)$ is the least integer $w$ such that for all
$r\in V(R)$, $|W_r\cup \bigcup_{e\sim r}X_e|\leq w+1$, where $e\sim
r$ means that $r$ is head or tail of $e$. The {\em directed
tree-width} of $D$, ${\rm dtw}(D)$, is the least integer $w$ such
that $D$ has an arboreal decomposition of width $w$.

The following  lemma is well-known
\cite{barat,berwangerSTACS06,johnsonJCT82,obdrzalekSODA06} and easy
to prove using just the definitions above.

\begin{lemma}\label{dw=0}
Let $D$ be a digraph. For  ${\rm d} \in \{{\rm dag}, {\rm dt}, {\rm
dp}\}$, we have ${\rm dw}(D)=0$ if and only if $D$ is acyclic.
\end{lemma}

\begin{lemma}\label{dpwbound}
For a digraph $D$, we have ${\rm dtw}(D)\le {\rm dpw}(D).$
\end{lemma}
\pf Let $Y_1,Y_2,\ldots, Y_k$ be the bags in a DPD
of $D$. We may assume that all bags are distinct. Define an arboreal
decomposition of $D$, where the arborescence is the directed path
$12\ldots k$, as follows: $W_1=Y_1$, $W_i=Y_i\setminus Y_{i-1}$ for
each $i=2,3,\ldots ,k$ and if $e=(i,i+1)$ we let $X_e=Y_i \cap
Y_{i+1}$. This arboreal decomposition is of the same width as the
DPD and we are done. \qed

\2

One of the main algorithmic results in \cite{johnsonJCT82} is on the
following linkage problem. Let $$\sigma=(s_1,t_1,s_2,t_2,\ldots
,s_p,t_p)$$ be a sequence of $2p$ vertices of a digraph $D$,
(vertices in $\sigma$ are not necessarily distinct). A 
{\em hamiltonian $\sigma$-linkage} of $D$ is a collection of 
$p$ directed paths $P_1,P_2,\ldots ,P_p$ such that 
$V(P_1)\cup \ldots \cup V(P_p)=V(D)$, $P_i$ starts at $s_i$ and 
terminates at $t_i$, $1\le i\le p,$ and
$(V(P_i)\setminus \{s_i,t_i\})\cap (V(P_j)\setminus \{s_j,t_j\})=\emptyset$ 
for all $1\le i<j\le p$. In the {\em hamiltonian
linkage problem}, given $\sigma$ we are to check whether there is a
hamiltonian $\sigma$-linkage of $D$.

\begin{theorem} \cite{johnsonJCT82}\label{dtwth}
For every fixed positive integer $p$ and every fixed nonnegative integer
$w$ the hamiltonian linkage problem with input sequence $\sigma$ of $2p$
vertices for digraphs of directed tree-width at most $w$ is
polynomial time solvable.
\end{theorem}

\section{New Results on MinLOB}

If $P$ is a directed path and vertices $a,b$ are, in that
order, on $P$, then we denote the $a-b$-segment of $P$ by $P[a,b]$,
and by $P[b,*]$ we mean the $b-t$-segment of $P$, where $t$ is the
terminal vertex of $P$.

\begin{theorem}
MinLOB is NP-hard for digraphs of directed path-width (directed
tree-width, DAG-width, respectively) 1.
\end{theorem}
\pf We prove the theorem by reduction of 3SAT to MinLOB. We use the
following gadget $H$, the digraph with vertex set
$V(H)=\{x_1,y_1,z_1,x_2,y_2,z_2\}$ and arc set
 $A(H) = \{ x_1y_1, y_1z_1, z_1x_1, x_1x_2, y_1y_2, z_1z_2, x_2z_2,
 z_2y_2, y_2x_2\}$.
It is easy to verify that $H$ has the following properties: \\[1mm]
(i) there exists a hamiltonian $(x_1,x_2)$-linkage $P_x$ of $H$, \\
(ii) there exists a hamiltonian $(x_1,x_2,y_1,y_2)$-linkage of $H$, \\
(iii) there exists an hamiltonian $(x_1,x_2,y_1,y_2,z_1,z_2)$-linkage of $H$, \\
 (iv) if $P_x$ is a hamilton path  of $H$ starting at $x_1$ then $P_x$
 ends in $x_2$, \\
 (v) if $P_x$ and $P_y$ are vertex disjoint paths in $H$ starting
 at $x_1$ and $y_1$, respectively, which go through all vertices of
 $H$, then either $P_ x$ ends in $x_2$ and $P_y$ ends in $y_2$, or $P_x$
 ends in $y_2$ and $P_y$ ends in $z_1$.  \\[1mm]
Analogous statements hold for each permutation of $x,y,z$.

Consider an instance $I$ of 3SAT with variables
$v^1,v^2,\ldots,v^k$ and clauses $C_1,C_2,\ldots,C_p$. Construct a
digraph $D=D(I)$ as follows: For each clause $C_j$ let $H_j$ be a
copy of $H$. If $C=\alpha + \beta + \gamma$, where $\alpha, \beta$,
and $\gamma$ are literals, denote the vertices of $H_j$ by
$\alpha_1(H_j), \beta_1(H_j), \gamma_1(H_j), \alpha_2(H_j),
\beta_2(H_j), \gamma_2(H_j)$. (Occasionally, when we do not wish to
specify the variables $\alpha, \beta, \gamma$, we denote the vertices
simply by $x_1(H_j),\ldots,z_2(H_j)$.) We also introduce a vertex $u_i$ for
each variable $v^i$ and a root vertex $r$. So
\[ V(D) = \{ r, u_1,u_2,\ldots,u_k \} \cup \bigcup_{j=1}^p V(H_j),
\]
and $D$ is a graph of order $6p+k+1$. \\
The arc set of $D$ consists of $\bigcup_{j=1}^p E(H_j)$, arcs $ru_i$
for $i=1,2,\ldots,k$ and the arcs in the sets
 ${\rm Arc}(v^1), {\rm Arc}(\overline{v^1}),\ldots, {\rm Arc}(v^k),
 {\rm Arc}(\overline{v^k})$
defined as follows. Consider a variable $v^i$. Let
$C_{j_1},C_{j_2},\ldots,C_{j_s}$, with $j_1<j_2<\ldots<j_s$, be the
clauses containing $v^i$ as literal. Then the set ${\rm Arc}(v^i)$
contains the arcs  $u_i v^i_1(H_{j_1})$,
 $v^i_2(H_{j_1}) v^i_1(H_{j_2})$,
 $v^i_2(H_{j_2}) v^i_1(H_{j_3})$, \ldots,
 $v^i_2(H_{j_{s-1}}) v^i_1(H_{j_s})$. Similarly let $C_{h_1},C_{h_2},\ldots,C_{h_t}$, with
$h_1<h_2<\ldots<h_t$, be the clauses containing $\overline{v^i}$ as
literal. Then the set ${\rm Arc}(\overline{v^i})$ contains the arcs
 $u_i \overline{v^i}_1(H_{h_1})$,
 $\overline{v^i}_2(H_{h_1}) \overline{v^i}_1(H_{h_2})$,
 $\overline{v^i}_2(H_{h_2}) \overline{v^i}_1(H_{h_3})$,\ldots,
 $\overline{v^i}_2(H_{h_{t}-1}) \overline{v^i}_1(H_{h_t})$.
This completes the construction of $D$.

We prove that \begin{equation}\label{dw=1} {\rm dtw}(D)={\rm
dagw}(D)={\rm dpw}(D)=1 \end{equation} Since $D$ is not acyclic, by
Lemma \ref{dw=0}, every width parameter in (\ref{dw=1}) is positive
and, by Lemma \ref{dpwbound}, it is enough to show that ${\rm
dpw}(D)\leq 1.$ It can be easily checked that the following bags form
a DPD of $D$ of width $1$:
\[ \{r\}, \{u_1\}, \{u_2\},\ldots,\{u_k\},  \]
\[  \{z_1(H_1),y_1(H_1)\},
    \{y_1(H_1),x_1(H_1)\}, \{x_2(H_1),y_2(H_1)\}, \{y_2(H_1),z_2(H_1)\},   \]
\[  \ldots, \{z_1(H_p),y_1(H_p)\},
    \{y_1(H_p),x_1(H_p)\}, \{x_2(H_p),y_2(H_p)\}, \{y_2(H_p),z_2(H_p). \]
We now show that $D$ has an out-branching with exactly $k$ leaves if and
only if $I$ is satisfiable.

Given a valid truth assignment to $v^1,\ldots,v^k$ we construct an
out-branching $B$ of $D$ with $k$ leaves as follows. Root $B$ at
$r$. Let $ru_1, ru_2,\ldots, ru_k \in E(B)$. If variable $v^i$ has
truth value TRUE then add all arcs in ${\rm Arc}(v^i)$ to $A(B)$.
Then these arcs, together with suitably (i.e., according to
properties (i), (ii) and (iii) of $H$) chosen
$v^i_1(H_j)-v^i_2(H_j)$ paths through those $H_j$ which correspond
to the $C_j$ containing $v^i$ as a literal, yield a path $P(v^i)$
starting at $u_i$. Similarly, if variable $v^i$ has truth value
FALSE then add all arcs in ${\rm Arc}(\overline{v^i})$ and suitably
chosen $\overline{v^i}_1(H_j)-v^i_2(H_j)$ paths to $A(B)$ and obtain
a path $P(\overline{v^i})$ starting at $u_i$. Since these $k$ paths,
attached to the vertices $u_1,\ldots,u_k$, go through all vertices
in  $V(D)$, $B$ is an out-branching of $D$ with exactly $k$ leaves.

Given an out-branching $B$ with exactly $k$ leaves of $D$, we derive
an assignment of truth values to the variables $v^1,\ldots,v^k$ that
satisfies each clause $C_j$ and thus $I$. We note that $B$ must be
rooted at $r$ since ${\rm d}_D^-(r)=0$ and that $ru_i\in A(B)$ for
$i=1,2,\ldots,k$ since ${\rm d}_D^-(u_i)=1$. So ${\rm d}_T^+(r)=k$,
hence the subtree of $T$ rooted at $u_i$ is a path $P_i$ for
$i=1,2,\ldots,k$.

Consider a subgraph $H_j$ of $D$. A path $P_i$ that intersects with
$H_j$ is said to be $H_j$-compatible if $P_i$ enters $H_j$ at $x_1$
and leaves at $x_2$, or it enters $H_j$ at $y_1$ and leaves at
$y_2$, or it enters $H_j$ at $z_1$ and leaves at $z_2$. We now show
that $B$ can be modified, without changing the number of leaves, so
that whenever a path $P_i$ and a gadget $H_j$ intersect, $P_i$ is
$H_j$-compatible. Consider a fixed $H_j$. First assume that $P_i$ is
the only path that intersects $H_j$. By property (iv) $P_i$ is
$H_j$-compatible. Next assume that two paths, $P_h$ and $P_i$ say,
intersect $H_j$ and that they enter $H_j$ in, say, $x_1$ and $y_1$,
respectively. By property (v) either $P_h$ and $P_i$ are
$H_j$-compatible, or $P_i$ ends in $z_1$ and $P_h$ ends in $y_2$. In
the latter case let $P_h'$ be the union of $P_h[u_h,x_1]$ and the
path $x_1,x_2$, and let $P_i'$ be the union of $P_i[u_i,y_1]$, the
path $y_1,z_1,z_2,y_2$ and $P_h[y_2,*]$, and replace $P_h$ and
$P_i$ by $P_h'$ and $P_i'$. Finally assume that three paths
$P_g,P_h,P_i$ intersect $H_j$. Then a similar construction yields
$H_j$-compatible paths $P_g', P_h'$ and $P_i'$. Clearly, replacing
$P_g, P_h, P_i$ by $P_g', P_h', P_i'$ if necessary does not change
the number of leaves of $B$, nor does it create any
incompatibilities. Hence repeating this step for all $H_j$
eventually yields an out-branching in which every path $P_i$ that
intersects a gadget $H_j$ is $H_j$-compatible.

Note that vertex $u_i$ has two out-neighbors in $D$, $v^i_1(H_{j_1})$ and
$\overline{v^i}_1(H_{h_1})$, where $C_{j_1}$ ($C_{h_1}$) is the
first clause to contain $v^i$ ($\overline{v^i}$) as a literal, and that $T$
contains at most one of these arcs.  If
the first arc of $P_i$ is $u_iv^i_1(H_{j_1})$ then we assign the
value TRUE to $v^i$, if the first arc of $P_i$ is
$u_i\overline{v^i}_1(H_{h_1})$ then we assign the value FALSE to
$v^i$, and if $P_i$ has no arc we assign an arbitrary truth value to
$v^i$. It remains to show that this satisfies $I$.

Fix an arbitrary clause $C_j$ and consider $H_j$. There is at least
one path $P_i$ of the out-branching $B$ that intersects with $H_j$.
Assume that the first arc of $P_i$ is, say, $u_iv^i_1(H_{j_1})$ (for
$u_i\overline{v^i}_1(H_{h_1})$ the proof is analogous) and that $P$
passes through $H_{j_1}, H_{j_2},\ldots$ before reaching $H_j$.
Since $P_i$ is compatible with $H_{j_1}, H_{j_2},\ldots,H_j$, it
enters $H_{j_1},H_{j_2},\ldots,H_j$ in $v^i_1(H_{j_1}),
v^i_1(H_{j_2}),\ldots,v^i_1(H_j)$. Hence clauses $C_{j_1},
C_{j_2},\ldots,C_j$ contain $v^i$ as a literal. But since we
assigned the value TRUE to $v^i$, clause $C_j$ is satisfied. Since
$C_j$ was arbitrary, all clauses and thus $I$ are satisfied. \qed

\begin{theorem}\label{posth} Let ${\rm d} \in \{{\rm dag},
{\rm dt}, {\rm dp}\}$. For every fixed positive integer $k$ and
every fixed nonnegative integer $w$, we can check, in polynomial
time, whether a digraph $D$ with ${\rm dw}(D)\le w$ has an
out-branching with at most $k$ leaves.
\end{theorem}
\pf Let $D$ be a digraph. By Lemma \ref{dpwbound}, if ${\rm
dpw}(D)\le k$ then ${\rm dtw}(D)\le k.$ It is shown in
\cite{berwangerSTACS06} that if ${\rm dagw}(D)\le k$ then ${\rm
dtw}(D)\le 3k+1.$

Thus, we may assume that $D$ is of directed tree-width at most $w$,
for some integer $w$, and let $B$ be an out-branching in $D$ with at
most $k$ leaves. Let $X(B)$ be the set consisting of the root, the
leaves and the branching vertices of $B$. It is not difficult to
show that $|X(B)|\le 2k.$ Now contract each directed path of $B$
between two vertices of $X(B)$ into an arc (between the vertices of
$X(B)$) and observe that we have obtained an out-tree $B'$ with
exactly $|X(B)|$ vertices. We call $B'$ the {\em contraction} of
$B$.

Now let $Y\subseteq V(D)$, $|Y|\le 2k$, and let $T$ be an
out-branching in $D[Y]$ with arcs $A(T)=\{(s_1,t_1),(s_2,t_2),\ldots
,(s_{|Y|-1},t_{|Y|-1})\}$. Using the algorithm of Theorem
\ref{dtwth} with input $(s_1,t_1,s_2,t_2,\ldots
,s_{|Y|-1},t_{|Y|-1})$, we can check, in polynomial time, whether
$D$ contains an out-branching $B^*$ whose contraction is $T$.

Thus, to find an out-branching in $D$ with minimum number of leaves,
we can use the following procedure. We generate all subsets of
$V(D)$ with at most $2k$ vertices and, for each such subset $Y$, we
generate all out-branchings $T$ in $D[Y]$. For each $T$ we use the
algorithm of Theorem \ref{dtwth} to verify whether $D$ has an
out-branching whose contraction is $T$. Finally, we find a minimum
leaf out-branching among all the outputs of the algorithm.

Observe that for each $Y$, by Cayley's formula on the number of
spanning trees in a complete graph, there are at most $|Y|^{|Y|-1}$
out-branchings of $D[Y]$ and that there are less than $|V(D)|^{2k+1}$
sets $Y$ with $|Y|\le 2k$. Thus, in our procedure, we use the
algorithm of Theorem \ref{dtwth} less than $|V(D)|^{2k+1}\cdot
(2k)^{2k-1}$ times, which shows that the running time of the
procedure is polynomial.\qed

\2

{\bf Acknowledgements} Research of Gutin and Kim was supported by EPSRC.


\begin{thebibliography}{99}


\bibitem{bang2000} J. Bang-Jensen and G. Gutin, {\em Digraphs: Theory, Algorithms
and Applications}, Springer, 2000; freely available online at {\tt
www.cs.rhul.ac.uk/books/dbook/}

\bibitem{barat} J. Bar\'at, Directed
path-width and monotonicity in digraph searching, {\em Graphs and
Combinatorics} 22 (2006), 161--172.

\bibitem{berwangerSTACS06} D. Berwanger, A. Dawar, P. Hunter and S. Kreutzer, DAG-width and
parity games, Proc. 23rd Annual Symp. on Theoretical Aspects of
Computer Science (STACS), 2006, 524--536.

\bibitem{demers2000} A. Demers and A. Downing, Minimum leaf spanning
tree, US Patent no. 6,105,018, August 2000.

\bibitem{gutinAAIM08} G. Gutin, I. Razgon and E.J. Kim, Minimum Leaf
Out-Branching Problems, {\em Proc. AAIM'08}, Lect. Notes Comp. Sci.
5034 (2008), 235--246.

\bibitem{hunterSODA07} P. Hunter and S. Kreutzer, Digraph Measures: Kelly Decompositions,
Games, and Orderings, Proc.  18th ACM-SIAM Symp. on Discrete
Algorithms (SODA), 2007, 637--644.

\bibitem{johnsonJCT82} T. Johnson, N.
Robertson, P.D. Seymour and R. Thomas, Directed Tree-Width, {\em J.
Comb. Theory, Ser. B} 82 (2001), 138--154.

\bibitem{obdrzalekSODA06}  J. Obdrzalek, DAG-width - Connectivity Measure for Directed Graphs,
Proc. 17th Annual ACM-SIAM Symp. on Discrete algorithms (SODA),
2006, 814 -- 821.





\end{thebibliography}
\end{document}